\documentclass[preprint]{imsart}
\usepackage{DEF}
\usepackage{DEF-script}
\usepackage{graphicx}

\startlocaldefs

\def\dx{\d x}
\def\dy{\d y}

\def\mch{Markov chain}

\def\hthsn{\hat\th_{T_{R(n)}}}
\def\fc{{\bar f}}
\def\asvar{\sigma_{\rm as}^{2}(f)}
\def\n0{n_{0}}
\def\Vu{\Vert_{V}}
\endlocaldefs

\begin{document}

\begin{frontmatter}

\title{Nonasymptotic bounds on the estimation error
for regenerative MCMC algorithms
\protect\thanksref{T1}}
\runtitle{Regenerative MCMC}
\thankstext{T1}{Work partially supported by Polish Ministry of Science and Higher Education  Grant No. N N201387234.}

\begin{aug}
\author{\fnms{Krzysztof} \snm{{\L}atuszy\'{n}ski}, 
\ead[label=e1]{latuch@gmail.com}}
\author{\fnms{B{\l}a\.zej} \snm{Miasojedow}
\ead[label=e2]{bmia@mimuw.edu.pl}}
\and
\author{\fnms{Wojciech} \snm{Niemiro}\corref{}
\ead[label=e3]{wniem@mat.uni.torun.pl}
\ead[label=u1,url]{http://www2.warwick.ac.uk/fac/sci/statistics/staff/research/latuszynski/}}

\runauthor{K. {\L}atuszy\'nski et al.}

\affiliation{University of Warwick, University of Warsaw and Nicolaus Copernicus University}

\address{K. {\L}atuszy\'nski\\ Department of Statistics\\
University of Warwick\\ CV4 7AL, Coventry, UK\\
\printead{e1}\\
\printead{u1}}

\address{B. Miasojedow\\Institute of Applied Mathematics and Mechanics\\ University of Warsaw\\ 
Banacha 2, 02-097 Warszawa, Poland\\
\printead{e2}}

\address{W. Niemiro\\Faculty of Mathematics and Computer Science\\ Nicolaus Copernicus University\\ 
Chopina 12/18, 87-100 Toru\'{n}, Poland\\
\printead{e3}}
\end{aug}

\begin{abstract} MCMC methods are used in Bayesian statistics not only to sample from posterior distributions but also to estimate expectations. Underlying functions are most often defined on a continuous state space and can be unbounded. 
We consider a regenerative setting and Monte Carlo estimators based on i.i.d.\ blocks of a Markov chain trajectory. The main result is an inequality for the mean square error. We also consider confidence bounds. We first derive the results in terms of the asymptotic variance and then bound the asymptotic variance for both uniformly ergodic and geometrically ergodic Markov chains.
\end{abstract}



\end{frontmatter}

\section{Introduction}\label{Sec: Intro}

Suppose that we are to estimate the expectation of a function, possibly unbounded and defined on a high dimensional space, with respect to some probability density which is known only up to a normalising constant. Such problems arise in Bayesian inference and are often solved
using Markov chain Monte Carlo (MCMC) methods. The idea is to simulate a Markov chain 
converging to the target distribution and take ergodic averages as estimates of the expectation.
It is essential to have explicit and reliable bounds which provide information about how long
the algorithms must be run to achieve a prescribed level of accuracy (c.f. \cite{Ros95JASA, HoJo_honest, CaHaJoNe}).

We consider MCMC algorithms which use independent and identically distributed 
random blocks of the underlying Markov chain, each block starting and ending at 
consecutive regeneration times. In fact, we propose a sequential version of regenerative estimator, for which the length of trajectory is ``nearly fixed''. This methodology is a promising alternative to both
fixing the total length of the trajectory and fixing the number of regeneration cycles \cite{Myk95, Ros95JASA, HoJoPrRos, BeCle1, BeCle2, BeCle3}. 
The  simulation scheme is easy to implement, provided that the regeneration times can be identified.
We introduce our estimator and discuss its properties in Section \ref{Sec: MainTh}.


The regenerative/sequential simulation scheme which we propose allows us to use directly the tools of the renewal theory and statistical sequential analysis. Our goal is to obtain quantitative bounds on the error of MCMC estimation. We aim at explicit nonasymptotic results. To this end we split the analysis into independent parts. 

First, in Section \ref{Sec: MainTh}, we derive inequalities on the mean square error (MSE) in terms of the asymptotic variance of the chain. This is obtained under very weak assumptions. We require a one step minorization condition (Assumption \ref{as: SmallSetCond}) and an integrability conditions that are essentially equivalent to those needed for Central Limit Theorems (CLTs) for nontrivial target functions.
The proof of our main result, \mbox{Theorem \ref{th: BasicMSE},} depends on a classical result of Lorden \cite{Lord} about the mean ``excess time'' for renewal processes and also on the two Wald's identities.

Next, in Section \ref{Sec: Conf}, we consider confidence estimation via a median trick that leads to an exponential inequality and we argue that our nonasymptotic bounds are not far off the asymptotic approximation based on the CLT.

Finally, we proceed to express the bounds in terms of computable quantities. In Section \ref{Sec:asV_unif} we consider uniformly ergodic chains, where bounding the asymptotic variance is straightforward. Moreover, in case of a bounded target function we can compare our approach to the well known exponential inequalities for Doeblin chains. Our bound is always within a factor of at most $40\beta$ of the exponential inequality (where $\beta$ is the regeneration parameter of the Doeblin chain), so it will turn out sharper for many examples of practical interest, where $\beta$ is small. In \mbox{Section \ref{Sec: asVar}} we assume the most general setting that motivates our work, namely a drift towards a small set, to replace the unknown asymptotic variance by known drift parameters. Our Assumption \ref{as: DriftCond} is quite similar to many analogous drift conditions
known in the literature, see e.g.\ \cite{MeynTw, RoTw, Bax}. For aperiodic chains 
Assumption \ref{as: DriftCond} implies geometric ergodicity, but we do not need
aperiodicity for our purposes. We build on some auxiliary results of \cite{Bax} to derive bounds on the asymptotic variance.

The nonasymptotic confidence intervals we derive are valid in particular for unbounded target functions and Markov chains that are not uniformly ergodic. Our assumptions are comparable (in some cases identical) to those required for asymptotically valid confidence intervals (c.f. \cite{CaHaJoNe, BeLa_asVar, FleJo, BeCle2}).  Moreover the bounds are expressed in terms of known quantities and thus can be of interest for MCMC practitioners. In \mbox{Section \ref{Sec: Discuss}} we discuss connections with related results in literature from both applied and theoretical viewpoint.  

One of the benchmarks for development of MCMC technology is the important hierarchical Bayesian model of variance components \cite{Ros95, HoGe98, JoHo}, used e.g.\ for small area estimation in survey sampling and in actuarial mathematics. We illustrate our theoretical results 
with a simple example which can be regarded as a part of this model. Since the analytic solution is known in this example, it is possible to assess the tightness of our bounds.
The full model of variance components will be considered in \cite{LaNieVarComp} and \cite{LaNie_nonas}.
\goodbreak

\section{Regenerative simulation}\label{Sec: Dreg}

Let $\pi$ be a probability distribution on a Polish space $\X$. Consider
a Markov transition kernel $P$ such that $\pi P=\pi$, 
that is $\pi$ is stationary \wrt  $P$. Assume $P$ is $\pi$-irreducible. 
The regeneration/split construction of Nummelin \cite{Num78} and Athreya 
and Ney \cite{AthN78} rests on the following assumption.
\begin{ass}[Small Set]\label{as: SmallSetCond} 
There exist a Borel set $J\subseteq\X$ of positive $\pi$ measure, a number $\beta>0$ and 
a probability measure $\nu$ such that 
\begin{equation*}
\qquad
    P(x,\cdot)\geq \beta \Ind(x\in J)\nu(\cdot).
\end{equation*}
\end{ass}
Under Assumption \ref{as: SmallSetCond} we can define a bivariate {\mch}   $(X_n,\Gamma_n)$  
on the space $\X\times\{0,1\}$ in the following way. Variable $\Gamma_{n-1}$ depends only on $X_{n-1}$ via $\Pr(\Gamma_{n-1}=1|X_{n-1}=x)=\beta \Ind(x\in J)$. 
The rule of transition from $(X_{n-1},\Gamma_{n-1})$ to $X_{n}$ is given by 
\begin{equation*}
\qquad
\begin{split}
  &\Pr(X_{n}\in A|\Gamma_{n-1}=1,X_{n-1}=x)=\nu(A), \\
  &\Pr(X_{n}\in A|\Gamma_{n-1}=0,X_{n-1}=x)=Q(x,A), \\ 
\end{split}
\end{equation*}
where $Q$ is the normalized ``residual'' kernel given by
\begin{equation*}
\qquad
  Q(x,\cdot):=\frac{P(x,\cdot)- \beta \Ind(x\in J)\nu(\cdot)}{1-\beta \Ind(x\in J)}.
\end{equation*}
Whenever $\Gamma_{n-1}=1$, the chain regenerates at moment $n$.
The regeneration epochs are
\begin{equation*}
\begin{split}\qquad
   &T_1:=\min\{n\geq 1: \Gamma_{n-1}=1\},\\
   &T_k:=\min\{n\geq T_{k-1}: \Gamma_{n-1}=1\}.
\end{split}
\end{equation*}
Write $\tau_{k}=T_{k}-T_{k-1}$.  Unless specified otherwise, we assume that $X_{0}\sim\nu(\cdot)$ and therefore $T_{0}:=0$ is also a time of regeneration. 
Symbols $\Pr$ and $\Ex$ without subscripts will be shorthands for $\Pr_{\nu}$ and $\Ex_{\nu}$ while initial distributions other than $\nu$ will be explicitly indicated.
The random blocks
\begin{equation*}
\qquad
     \Xi_k :=(X_{T_{k-1}},\ldots,X_{T_{k}-1},\tau_{k})\\
\end{equation*}
for $k=1,2,3,\ldots$ are \iid  

We assume that we can
simulate the split chain $(X_n,\Gamma_n)$, starting from $X_{0}\sim\nu(\cdot)$. Put differently, 
we are able to \textit{identify} regeneration times $T_{k}$. 
Mykland et al.\ pointed out in \cite{Myk95} that actual sampling from $Q$ can be avoided. Assume that the chain $X_n$ is generated using transition probabability $P$. Let
$\nu(\dy)/P(x,\dy)$ denote the Radon-Nikodym derivative (in practice, the ratio of densities).
Then we can recover the regeneration indicators via
\begin{equation*}
\qquad
   \Gamma_{n-1}=\Ind\left\{U_{n}< 
        \Ind(X_{n-1}\in J)\frac{\beta \nu(\d X_{n})}{P(X_{n-1},\d X_{n})}\right\},
\end{equation*}
where $U_{n}$ is a sequence of \iid\ uniform variates independent of $X_{n}$. 
If sampling from the renewal distribution $\nu(\cdot)$ is difficult then we can start the
simulation from an arbitrary state, discard the initial part of the trajectory
before the first time of regeneration and consider only blocks  $\Xi_{k}$ for $k=2,3,\ldots$,
that is begin at $T_1$ instead of $T_0=0$. Thus in the regenerative scheme there is a very precise recipe for an  ``absolutely sufficient burn-in'' time.

\section{Main Theorem}\label{Sec: MainTh}

Let $f:\X\to\Rl$ be a Borel function. The objective is to compute 
(estimate) the quantity
\begin{equation*}
\qquad
  \th:=\pi(f)=\int_\X \pi(\dx)f(x).
\end{equation*}
We assume that $\th$ exists, i.e.\ $\pi(|f|)<\infty$. 
Regenerative estimators of $\th$ are based on the block sums
\begin{equation*}
\qquad
     \Xi_{k}(f) :=\sum_{i=T_{k-1}}^{T_{k}-1} f(X_i).
\end{equation*}
Let us now introduce a sequential version of regenerative estimator. Fix $n$ and define
\begin{equation}\label{eq: R}
\qquad
      R(n):=\min\{r: T_{r}\geq n\}.
\end{equation}
Our basic estimator is defined as follows.
\begin{equation}\label{eq: est}
\qquad
   \hthsn := \frac{1}{T_{R(n)}} \sum_{i=1}^{R(n)}\Xi_{k}(f) = \frac{1}{T_{R(n)}}\sum_{i=0}^{T_{R(n)}-1}f(X_{i}). 
\end{equation}
In words: we stop simulation \textit{at the first moment of regeneration past $n$} and
compute the usual sample average. Note that we thus generate a random number of blocks.
Our regenerative scheme requires only as many blocks as necessary to make the length of trajectory at least $n$  and the ``excess time'' $T_{R(n)}-n$ will be shown to be small compared to $n$. 
\goodbreak

The result below bounds the mean square error (MSE) of the estimator defined
by \eqref{eq: est}, \eqref{eq: R} and the expected number of samples used to compute it. 
Let $\fc:=f-\pi(f)$. 

\begin{thm}\label{th: BasicMSE}
If Assumption \ref{as: SmallSetCond} holds, $\Ex (\Xi_{1}(\fc))^{2}<\infty$ and 
$\Ex \tau_{1}^{2}<\infty$ then
\begin{equation*}
(i)\quad
   \Ex\,(\hthsn-\th)^2\leq \frac{\asvar}{n^2}\,\Ex\,T_{R(n)}
\end{equation*}
and
\begin{equation*}
(ii)\quad
     \Ex\, T_{R(n)}\leq n + \n0,
\end{equation*}
where
\begin{equation*}
\qquad
    \asvar:=\frac{\Ex (\Xi_{1}(\fc))^{2}}{\Ex \tau_{1}} ,\qquad
     \n0:=\frac{\Ex \tau_{1}^{2}}{\Ex \tau_{1}}-1.
\end{equation*}
\end{thm}
\begin{cor}\label{cor: BasicMSE}
Under the same assumptions,
\begin{equation*}
\qquad
   \Ex\,(\hthsn-\th)^2\leq
      \frac{\asvar}{n}\left(1+\frac{\n0}{n}\right).
\end{equation*}
\end{cor}
Note that the leading term ${\asvar}/{n}$ in Corollary \ref{cor: BasicMSE} is
``asymptotically correct'' in the sense that, under our assumptions,
\begin{equation*}
\qquad
   \limn n\Ex\left(\frac{1}{n}\sum_{i=1}^{n}f(X_{i})-\th\right)^2 ={\asvar}\; \text { and }\; \limn
                                                             \frac{\Ex T_{R(n)}}{n}=1. 
\end{equation*}

\begin{rem}
Under Assumption \ref{as: SmallSetCond}, finiteness of $\Ex (\Xi(\fc))^{2}$ is a sufficient and necessary condition for the CLT to hold for Markov chain $X_{n}$ and function $f$. This fact is proved in \cite{BeLaLa} in a more general setting. For our purposes it is important to note 
that $\asvar$ in Theorem \ref{th: BasicMSE} is indeed the \textit{asymptotic variance} which
appears in the CLT. Constant $n_{0}$ bounds the ,,\textit{mean overshoot}'' or excess length of simulations over $n$.

\end{rem}

\begin{proof}[Proof of Theorem \ref{th: BasicMSE} (\textit{i})]
Note that
\begin{equation*}
\qquad
   \hthsn-\th=\frac{\sum\limits_{k=1}^{R(n)}\Xi_{k}(f)}{\sum\limits_{k=1}^{R(n)}\tau_{k}}-\th
             =\frac{1}{T_{R(n)}}\sum_{k=1}^{R(n)}d_{k},
\end{equation*}
where $d_k:=\Xi_{k}(f)-\th \tau_{k}=\Xi_{k}(\fc)$.
By the Kac theorem (\cite{MeynTw} or \cite{MCMCists}) we have
\begin{equation*}
\qquad
    \Ex\, \Xi_{k}(f)= m \pi(f)= m \th,
\end{equation*} 
where
\begin{equation*}
\qquad
  m:=\Ex \tau_{k}=\frac{1}{\beta\pi (J)}.
\end{equation*} 
Consequently the pairs $(d_{k},\tau_{k})$ are \iid\ with $\Ex d_{k}=0$ and $\var d_{k}=m\asvar$.
Since  $T_{R(n)}\geq n$, it follows that
\begin{equation*}
\qquad
   \Ex\,(\hthsn-\th)^2\leq 
         \frac{1}{n^{2}}\Ex\left(\sum_{k=1}^{R(n)}d_k\right)^2.
\end{equation*}
Since $R(n)$ is a stopping time with respect to
 $\G_k=\sigma((d_{1},\tau_{1}),\ldots,(d_{k},\tau_{k}))$, we are in a position to apply the 
two Wald's identities. The second identity yields
\begin{equation*}
\qquad
   \Ex\left(\sum_{k=1}^{R(n)}d_k\right)^2=\var\, d_{1}\,\Ex R(n)=
                   m\asvar\,\Ex R(n).
\end{equation*}
But in this expression we can replace $m\Ex R(n)$ by $\Ex T_{R(n)}$ because of the first Wald's identity:
\begin{equation*}
\qquad
   \Ex\, T_{R(n)}= \Ex \sum_{k=1}^{R(n)}\tau_k=\Ex\tau_{1}\,\Ex R(n)=m\Ex R(n)
\end{equation*}
and the claimed result follows. 
\end{proof}

We now focus attention on bounding the ``excess'' or ``overshoot'' time 
\begin{equation*}
\qquad \Delta(n):=T_{R(n)}-n.
\end{equation*}
To this end, let us recall a classical result of the (discrete time) renewal theory.
As before, when using symbols $\Pr$ and $\Ex$ without subscripts
we refer to the chain started at the renewal distribution $\nu$ and we write $m=\Ex \tau_{1}$. 
Let $\Delta(\infty)$ be a random variable having distribution 
\begin{equation*}
\qquad
    \Pr\,\left(\Delta(\infty)=i\right):=\frac{1}{m}\Pr(\tau_{1}>i)\quad\text{for } i=0,1,2,\ldots.
\end{equation*}
If the distribution of $\tau_{1}$ is aperiodic then it is well-known that 
$\Delta(n)\to\Delta(\infty)$ in distribution, as $n\to \infty$, but we will not
use this fact directly. Instead, we invoke the following elegant result.
\begin{prop}[Lorden \cite{Lord}]\label{pr: Lord}
\begin{equation*}
\qquad
    \Ex\,\Delta(n) \leq 2\, \Ex\,\Delta(\infty).
\end{equation*}
\end{prop}
For a newer simple proof of Lorden's inequality, we refer to \cite{Chang}. 
Proposition \ref{pr: Lord} gives us exacly what we need to conclude the proof 
of our main result.

\begin{proof}[Proof of Theorem \ref{th: BasicMSE} (\textit{ii})]
Write $p_{i}:=\Pr(\tau_{1} =i)$. We have
\begin{equation*}
\begin{split}
\qquad \Ex\,\Delta(\infty)& =\frac{1}{m} \sum_{i=1}^{\infty} i \sum_{j=i+1}^{\infty}p_j\\
                          & =\frac{1}{m} \sum_{j=2}^{\infty}p_{j}\sum_{i=1}^{j-1} i=
                                 \frac{1}{m}  \sum_{j=2}^{\infty}p_{j}\frac{j(j-1)}{2}\\
                      &=\frac{1}{m} \Ex\frac{\tau_{1} (\tau_{1} -1)}{2}\\
                      &=\frac{1}{2m}\Ex \tau_{1} ^2-\frac{1}{2}.
\end{split}
\end{equation*}
By the Lorden's theorem we obtain
\begin{equation*}
\qquad \Ex\,\Delta(n)\leq 2\,\Ex\,\Delta(\infty)\leq \frac{1}{m}\Ex \tau_{1} ^2-1,
\end{equation*}
which is just the desired conclusion.
\end{proof}

\section{Confidence estimation}\label{Sec: Conf}

Although the MSE is an important quantity in its own right, it can also
be used to  construct estimates with fixed precision at a given level of confidence. 
Suppose the goal is to obtain an estimator $\hat\th$ such that 
\begin{equation}\label{eq: conf}
  \qquad
 \Pr(|\hat\th-\th|>\eps)\leq \alpha,
 \end{equation}
for given $\eps>0$ and $\alpha>0$.
Corollary \ref{cor: BasicMSE} combined with the Chebyshev's inequality yields
the following bound:
\begin{equation}\label{eq: Chebysh}
\qquad
   \Pr\,(|\hthsn-\th|>\eps)\leq
     \frac{\asvar}{n\eps^{2}}\left(1+\frac{\n0}{n}\right).
\end{equation}
If $\alpha$ is small then
instead of using \eqref{eq: Chebysh} directly, 
it is better to apply the so-called ``median trick''. This is a method introduced in 1986 in  \cite{JeVaViz86}, later used in many papers concerned with computational complexity, eg.\  \cite{Gil, MIS} and further developed in \cite{MIS}. 
The idea is to compute the median of independent estimates
to boost the level of confidence. We simulate $l$ independent copies of the Markov chain:
\begin{equation*} 
\qquad
 X_{0}^{(j)},X_{1}^{(j)},\ldots,X_{n}^{(j)},\ldots\qquad(j=1,\ldots,l). 
\end{equation*}
Let $\hat\th^{(j)}$ be an estimator computed in $j$th repetition. 
The final estimate is $\hat\th:=\med(\hat{\th}^{(1)},\ldots,\hat{\th}^{(l)})$.
We require that $\Pr(|\hat\th^{(j)}-\th|>\eps)\leq \delta$ ($j=1,\ldots,l$) for some modest level
of confidence $1-\delta<1-\alpha$. This is ensured via Chebyshev's inequality.  The well-known Chernoff's bound gives for odd $l$,
\begin{equation}\label{eq: Chernoff}
\qquad 
\Pr\,(|\hat\th-\th|\geq \eps)\leq \frac{1}{2}\left[4\delta(1-\delta)\right]^{l/2}
=\frac{1}{2}\exp\left\{\frac{l}{2}\ln\left[4\delta(1-\delta)\right]\right\}.
\end{equation} 
In this way we obtain an exponential inequality for the probability of large deviations without requiring the underlying variables to be bounded or even to have a moment generating function. It is pointed out in \cite{MIS} that under some assumptions there is a universally optimal choice of $\delta$. More precisely, suppose that the bound on $\Pr(|\hat\th^{(j)}-\th|>\eps)$ is of the form $\const/n$ where $n$ is the sample size used in a single repetition. Then the overall number of samples $nl$ is the least if we choose
$\delta^{*}\approx  0.11969$. The details are described in \cite{MIS}. 
This method can be used in conjunction with our regenerative/sequential scheme.
The right hand side of \eqref{eq: Chebysh} \textit{approximately} behaves like $\const/n$. Therefore the following strategy is reasonably close to optimum. 
First choose $n$ such that the right hand side of \eqref{eq: Chebysh} is less than or equal to $\delta^{*}$. Then choose $l$ big enough  to
make the right hand side of \eqref{eq: Chernoff}, with $\delta=\delta^{*}$, less than or equal to $\alpha$. Compute estimator $\hthsn$ repeatedly, using $l$ independent runs of the chain.
We can easily see that \eqref{eq: conf} holds if 
\begin{equation*}
\begin{split}
\qquad 
 & n \geq \frac{C_1 \asvar}{\eps^{2}}+\n0,\\
 & l \geq C_2 \ln(2\alpha)^{-1} \;\text{ and $j$ is odd},
\end{split}
\end{equation*}
where $C_1:=1/\delta^{*}\approx  8.3549$ and $C_2:=2/{\ln\left[4\delta^{*}(1-\delta^{*})\right]^{-1}}\approx 2.3147$ are absolute constants.
By Theorem \ref{th: BasicMSE} (ii) the overall (expected) number of generated samples is
\begin{equation}\label{eq: final_1} 
\qquad
  \Ex T_{R(n)}l\sim nl\sim {C} \frac{\asvar}{\varepsilon^{2}}\log (2\alpha)^{-1},
\end{equation}
where $C=C_1C_2\approx  19.34$ and notation $\textrm{Left}(\alpha, \varepsilon) \sim \textrm{Right}(\alpha, \varepsilon) $ means that $\textrm{Left}/\textrm{Right} \to 1$ as $\alpha,\eps\to 0$.  
To see how tight are the bounds, let us compare \eqref{eq: final_1} with
the familiar {asymptotic approximation}, based on the CLT. We obtain
\begin{equation*}
\qquad 
  \lim_{\eps\to 0}\; \Pr(|\hat{\th}_{n}-\th|> \varepsilon)=\alpha,
\end{equation*}
for the number of samples
\begin{equation}\label{eq: asymp} 
 \qquad
  n\sim \frac{\asvar}{\varepsilon^{2}}\left[\Phi^{-1}(1-\alpha/2)\right]^{2},
\end{equation}
where $\hat{\th}_{n}$ is a simple average over $n$ Markov chain samples, $\Phi^{-1}$ is a quantile function of the standard normal distribution.
Taking into account the fact that 
\begin{equation*} 
 \qquad \left[\Phi^{-1}(1-\alpha/2)\right]^{2}\sim 2 \log (2\alpha)^{-1},
\qquad(\alpha\to 0),
\end{equation*}
we arrive at the following conclusion. The right hand side of \eqref{eq: final_1}
is bigger than \eqref{eq: asymp} roughly by a constant factor of about 10 
(for small $\eps$ and $\alpha$). The important difference is that \eqref{eq: final_1} 
is sufficient for an \textit{exact} confidence interval while \eqref{eq: asymp} only for an
\textit{asymptotic} one.
 
\section{Bounding the asymptotic variance I - uniformly ergodic chains}\label{Sec:asV_unif}

\def\q {h}

We are left with the task of bounding $\asvar$ and $\n0$,  which appear in Theorem
\ref{th: BasicMSE}, by some computable quantities in typical situations of interest. The most important setting for applications - that of a geometrically ergodic Markov chain and unbounded target function $f$ - is deferred to the next section. Here we start with uniformly ergodic chains, where a direct comparison of our approach to exponential inequalities \cite{GlyOr, KontoLastraMeyn} is possible. We focus on \cite{KontoLastraMeyn} which is tight in the sense that it reduces to the Hoeffding bound when specialised to the i.i.d.\ case.

Uniform ergodicity of a Markov chain is equivalent to 
\begin{equation}\label{eqn_unif_erg} \qquad
P^{\q}(x, \cdot) \geq \beta \nu(\cdot) \quad \textrm{for every} \quad x \in \mathcal{X} \quad \textrm{and some integer} \quad h\geq 1.
\end{equation}
We refer to \cite{RobRos} or \cite{MeynTw} for definitions of uniform and geometric ergodicity of Markov chains and further details related to these notions. 

In the rest of this section we assume that $\q = 1$ and hence (\ref{eqn_unif_erg}) reduces to Assumption \ref{as: SmallSetCond} with $J = \mathcal{X}.$ This is the typical situation in applications. If $\q > 1$ then $P^{\q}$ inherits the ergodic properties of $P$ and one can use it for sampling. However, we acknowledge that if $P^{\q}$ is used, the identification of regeneration times can be problematic since the term $P^{\q}(x, \d y)$ - needed to execute the Mykland et al. trick - will be typically intractable.

Computing $n_0$ in the setting of this Section is clearly trivial, since the overshoot is distributed as a geometric random variable with parameter $\beta.$ 

The problem of bounding the asymptotic variance under (\ref{eqn_unif_erg}) was considered in \cite{BeLaLa}. Using results of their Section 5 with $\q = 1$ and applying basic algebra we obtain  
\begin{equation}\label{eqn_as_var_unif} \quad
\asvar \leq \sigma^2 \left(1 + \frac{2}{1 - \sqrt{1-\beta}}\right) = \sigma^2 \left(1 + 2\frac{1+ \sqrt{1-\beta}}{\beta}\right) \leq 4\sigma^2/\beta, 
\end{equation}
where $\sigma^2 = \pi \bar{f}^2$ is the stationary variance.

With reversibility one can derive a better bound. An important class of reversible chains are Independence Metropolis-Hastings chains (see e.g. \cite{RobRos}) that are known to be uniformly ergodic if and only if the rejection probability $r(x)$ is uniformly bounded from 1 by say $1- \beta$. This is equivalent to the candidate distribution being bounded below by $ \beta \pi$ (c.f. \cite{MengersenTweedie, AtchadePerron}) and translates into (\ref{eqn_unif_erg}) with $\q = 1$ and $\nu(\cdot):=\pi(\cdot).$ In this setting and using reversibility Atchad\'e and Perron \cite{AtchadePerron} show that the spectrum of $P,$ say $\mathcal{S},$ is contained in $[0, 1-\beta].$ For the general case of reversible chains satisfying (\ref{eqn_unif_erg}) with $\q = 1$ results of \cite{RobRos_electr} lead to $\mathcal{S} \subseteq [-1+\beta, 1-\beta].$  By the spectral decomposition theorem for self adjoint operators (see e.g. \cite{Geyer, KipnisVaradhan}) in both cases we have
\begin{equation}\label{eqn_as_var_IMH} \qquad
\asvar \leq \int_{\mathcal{S}} \frac{1+s}{1-s}E_{f,P}(\d s) \leq \frac{2-\beta}{\beta} \sigma^2, 
\end{equation}
where $E_{f,P}$ is the spectral measure associated with $f$ and $P.$ The formula for $\asvar$ in (\ref{eqn_as_var_unif}) and (\ref{eqn_as_var_IMH}) depends on $\beta$ in an optimal way. Moreover (\ref{eqn_as_var_IMH}) is sharp. To see this consider the following example. 
\begin{exam}\label{example_sharp_as_var} Let $\beta \leq  1/2$ and define a Markov chain $(X_n)_{n\geq 0}$ on $\mathcal{X} = \{0,1\}$ with stationary distribution $\pi = \{1/2, 1/2\}$ and transition matrix \begin{equation}\nonumber \qquad P = \left[ \begin{array}{cc} 1-\beta/2 & \beta/2 \\ \beta/2 & 1-\beta/2
\end{array} \right]. \end{equation}
Hence $P = \beta \pi + (1-\beta) I_2$ and $P(x, \cdot) \geq \beta \pi.$
Moreover let $f(x) = x.$ Thus $\sigma^2 = 1/4.$ Now let us compute $\asvar .$
\begin{eqnarray*}
\asvar & = & \sigma^2 + 2 \sum_{i=1}^{\infty} \cov \{f(X_0), f(X_i)\} \\ &=& \sigma^2 + 2\sigma^2 \sum_{i=1}^{\infty} (1-\beta)^i =  \frac{2-\beta}{\beta} \sigma^2.
\end{eqnarray*}   
\end{exam}

To obtain an upper bound on the total simulation effort needed for $ \Pr(|\hat\th-\th|>\eps)\leq \alpha$ for our regenerative-sequential-median estimator $\hat\th,$ we now combine (\ref{eqn_as_var_unif}) and (\ref{eqn_as_var_IMH}) with (\ref{eq: final_1}) to obtain respectively 
\begin{equation}\label{eqn_unif_total_cost}\qquad
19.34\frac{4\sigma^2}{\beta \eps^2}\log(2\alpha)^{-1} \qquad \textrm{and} \qquad 19.34\frac{(2-\beta)\sigma^2}{\beta \eps^2}\log(2\alpha)^{-1}. 
\end{equation}
From Section \ref{Sec: Conf} and Example \ref{example_sharp_as_var} we conclude that in (\ref{eqn_unif_total_cost}) the form of functional dependence on all the parameters is optimal.

For $f$ bounded let $\|f\|_{\textrm{sp}} := \sup_{x\in \mathcal{X}} f(x) - \inf_{x \in \mathcal{X}} f(x)$ and consider the exponential inequality for uniformly ergodic chains from \cite{KontoLastraMeyn}. For the simple average over $n$ Markov chain samples, say $\hat\th_n$, for an arbitrary starting point $x,$ we have
\begin{equation}\nonumber \qquad
\Pr_{x}(|\hat\th_n-\th|>\eps) \leq 2 \exp \left\{ -\frac{n-1}{2}\left(\frac{2\beta}{\|f\|_{\textrm{sp}}}\eps - \frac{3}{n-1}\right)^2 \right\}.
\end{equation}
After identifying leading terms in the resulting bound for the simulation effort required for $ \Pr(|\hat\th_n-\th|>\eps)\leq \alpha$ and assuming, to facilitate comparisons, that $4\sigma^2 = \|f\|_{\textrm{sp}}^2,$ we see that
\begin{equation}\label{eqn_unif_exponential_cost}\qquad 
n \sim \frac{2\sigma^2}{\beta^2 \eps^2}\log(2\alpha)^{-1}.
\end{equation}

Comparing (\ref{eqn_unif_total_cost}) with (\ref{eqn_unif_exponential_cost}) yields a ratio of $40\beta$ or $20\beta$ respectively. This in particular indicates that the dependence on $\beta$ in \cite{GlyOr, KontoLastraMeyn} probably can be improved. We note that in examples of practical interest $\beta$ usually decays exponentially with dimension of $\mathcal{X}$ and our approach will often result in a lower total simulation cost. Moreover, in contrast to exponential inequalities of the classical form, our approach is valid for an unbounded target \mbox{function $f.$}

\section{Bounding the asymptotic variance II - drift condition}\label{Sec: asVar}

In this Section we bound $\asvar$ and $\n0$ appearing in Theorem
\ref{th: BasicMSE}, by computable quantities under drift condition and with possibly unbounded $f.$ Using drift conditions is a standard approach for establishing geometric ergodicity and our version is one of many equivalent drifts appearing in literature. Specifically, let $J$ be the small set which appears in Assumption \ref{as: SmallSetCond}. 
\begin{ass}[Drift]\label{as: DriftCond} 
There exist a function $V:\X\to [1,\infty[$, constants $\lambda<1$ and 
$K<\infty$ such that
\begin{equation*}
\qquad
    PV^{2}(x):=\int_{\X}P(x,\d y)V^2(y)\leq \begin{cases}
               \lambda^{2} V^{2}(x)& \text{for }x\not\in J,\\
               K^{2}& \text{for }x\in J,
              \end{cases}
\end{equation*}
\end{ass}
Unusual notation in the above drift condition  is chosen to simplify
further statements. Note that Assumption \ref{as: DriftCond} entails
\begin{equation}\label{eq: Drift}
\qquad
    PV(x)\leq \begin{cases}
               \lambda V(x)& \text{for }x\not\in J,\\
               K& \text{for }x\in J,
              \end{cases}
\end{equation}
because by Jensen's inequality $PV(x)\leq \sqrt{PV^2(x)}$. This simple observation is also exploited in 
\cite{LatPhD} and \cite{LaNie, LaNieVarComp}.
Assumptions \ref{as: SmallSetCond} and \ref{as: DriftCond} will allow us to derive explicit bounds on $\asvar$ and $\n0$ in terms of $\lambda$, $\beta$ and $K$, provided that function $\fc/V$ is bounded.


To simplify notation, let us write $T:=\min\{n\geq 1: \Gamma_{n-1}=1\}$  for the first time of regeneration and 
$\Xi:=\Xi_{1}$ for the first block. In contrast with the previous section, we will consider 
initial distributions of the chain different from $\nu$ and often equal to $\pi$, the
stationary measure. 
The following proposition appears e.g.\ in \cite{MCMCists} (for bounded $g$). The proof for nonnegative  $g$ is the same.
\begin{prop}\label{pr: SquareBlock}
For $g:\X\to [0,\infty[$, 
\begin{equation*}
\qquad           
       \Ex\,\Xi(g)^{2}=m \left[\Ex_\pi g(X_0)^2+2\sum_{n=1}^\infty \Ex_\pi
                           g(X_0)g(X_n)\Ind(T >n)\right].
\end{equation*}
\end{prop}

Our approach is based on the following result which is a slightly modified special case 
of Propositions 4.1 and 4.4 in \cite{Bax}, see also \cite{LT}. To make the paper reasonably self-contained we include the proof in Appendix \ref{sec_appendix_prop}.

\begin{prop}\label{pr: Bax} If \eqref{eq: Drift} holds, then
\begin{equation*}
\qquad           
      \Ex_x\sum_{n=1}^{T-1}V(X_n)\leq
              \frac{\lambda (V(x)-1)}{1-\lambda}+\frac{K-\lambda}{\beta(1-\lambda)}-1.
\end{equation*}
\end{prop}

Let us mention that a result similar to Theorem \ref{th: asVarBound} can also be obtained using
methods borrowed from \cite{DoMouRos}, c.f.\ also \cite{Ros02}, instead of \cite{Bax}. 
Although in the cited papers the inequalities are
derived for \textit{coupling}, they could easily be modified to work in the context of \textit{regeneration}.
We will not pursue this, because Proposition \ref{pr: Bax} is easier to apply.

The main result in this section is the following.
\begin{thm}\label{th: asVarBound} ~\begin{itemize} \item[(i)] Under Assumption \ref{as: SmallSetCond} and  \eqref{eq: Drift},
constant $\n0$ in Theorem \ref{th: BasicMSE} satisfies
\begin{equation*}
\quad 
    \n0 \leq 2\left[\frac{\lambda  \pi(V)-\lambda }{1-\lambda}
                             +\frac{K-\lambda}{\beta(1-\lambda)}-1\right].
\end{equation*}
\item[(ii)] If moreover 
$|\fc(x)| \leq V(x)$ then the asymptotic variance $\asvar$ satisfies
\begin{equation*}
\quad
   \asvar\leq \frac{1
+\lambda}{1-\lambda}\pi(V^2)+
2\left[\frac{K-\lambda-\beta}{\beta(1-\lambda)}\right]\pi(V).
\end{equation*}
\end{itemize}
\end{thm}

\begin{proof} (i) We apply Proposition \ref{pr: SquareBlock} to $g(x)=1$. Indeed, 
$\Ex T^{2}=\Ex \Xi(1)^{2}$ and 
\begin{equation*}
\begin{split}
\qquad 
  \Ex T^{2}/m &=\Ex \Xi(1)^2/m \\
              &\leq 1+2\Ex_\pi \sum_{n=1}^{T-1}V(X_n)\\
              &\leq 1+2\left[\lambda \frac{ \pi(V)-1}{1-\lambda}
                             +\frac{K-\lambda}{\beta(1-\lambda)}-1\right].
\end{split}
\end{equation*}
by Proposition \ref{pr: Bax}. The result follows because $\n0=\Ex T^{2}/m-1$.

(ii) By Proposition \ref{pr: SquareBlock} we have
\begin{equation*}
\begin{split}
\qquad 
  \asvar&=\Ex \Xi(\fc)^2/m\leq 
\Ex \Xi(V)^2/m\\
        &=\Ex_\pi V(X_0)^2+2\Ex_\pi\sum_{n=1}^{T-1}V(X_0)V(X_n)
\end{split}
\end{equation*}
We will use Proposition \ref{pr: Bax} to bound the second term.
 \begin{equation*}
\begin{split}
\qquad 
  &\Ex_\pi\sum_{n=1}^{T-1}V(X_0)V(X_n)=\Ex_\pi V(X_0)\Ex(\sum_{n=1}^{T-1}V(X_n)|X_0)\\
                   &=\int_{\X} \pi(\d x) V(x)\Ex_x\sum_{n=1}^{T-1}V(X_n)\\
                 &\leq \int_{\X} \pi(\d x) V(x)
                \left(\frac{\lambda (V(x)-1)}{1-\lambda}+\frac{K-\lambda}{\beta(1-\lambda)}-1\right)\\
    &=\frac{\lambda}{1-\lambda}\pi(V^2)+
\left[\frac{K-\lambda-\lambda\beta}{\beta(1-\lambda)}-1\right]\pi(V).
\end{split}
\end{equation*}
Putting everything together, we obtain
\begin{equation*}
\qquad
   \asvar\leq \pi(V^2)+
\frac{2\lambda}{1-\lambda}\pi(V^2)+
2\left[\frac{K-\lambda-\lambda\beta}{\beta(1-\lambda)}-1\right]\pi(V),
\end{equation*}
which is equivalent to the desired conclusion.
\end{proof}

Note that in Theorem \ref{th: asVarBound} we need only \eqref{eq: Drift}, that is the drift condition on $V$.
Asssumption \ref{as: DriftCond} is needed only to get a bound on $\pi(V^2)$. Indeed, it
implies that $\pi V^2=\pi PV^2\leq \lambda^2(\pi V^2-\pi(J)) +K^2 \pi(J)$, so
\begin{equation*}
\qquad
   \pi V^2\leq \pi(J) \frac{K^2 - \lambda^2}{1-\lambda^2}\leq\frac{K^2 - \lambda^2}{1-\lambda^2}.
\end{equation*}
Analogously, \eqref{eq: Drift} implies
\begin{equation*}
\qquad
   \pi V\leq \pi(J) \frac{K - \lambda}{1-\lambda}\leq\frac{K - \lambda}{1-\lambda}.
\end{equation*}
Our final estimates are therefore the following.
\begin{cor}\label{cor: asVarBound} ~ \begin{itemize}\item[(i)] Under Assumptions \ref{as: SmallSetCond} and  \ref{as: DriftCond},
\begin{equation*}
    \n0 \leq \frac{2}{(1-\lambda)\beta}
    \left[K \frac{1-\lambda(1-\beta)}{1-\lambda} -\beta \left(1 + \frac{\lambda^2}{1-\lambda}\right) -\lambda \right].
\end{equation*}

\item[(ii)] If $\Vert \fc \Vu:= \sup_{x}|\fc(x)|/V(x)<\infty$ then  

\begin{equation*}
   \asvar\leq \Vert \fc \Vu^{2}\frac{K^2(2+\beta) - 2K(2\lambda + \beta) +2\lambda^2 +2\lambda\beta - \lambda^2\beta}{(1-\lambda)^{2}\beta}.
\end{equation*}

\item[(iii)] Moreover, $\Vert \fc \Vu$ can be related to $\Vert f \Vu$ by

\begin{equation*}
\Vert \fc \Vu \leq \Vert f \Vu + \frac{\pi(J)(K-\lambda)}{(1-\lambda) \inf_{x \in \mathcal{X}}V(x)} \leq  \Vert f \Vu + \frac{K-\lambda}{1-\lambda}. 
\end{equation*}

\end{itemize}
\end{cor}

\begin{proof} To prove (iii) we compute
\begin{eqnarray*}
\Vert \fc \Vu & = & \sup_{x \in \mathcal{X}} \frac{|f(x) - \pi f|}{V(x)} \leq  \sup_{x \in \mathcal{X}} \frac{|f(x)| + |\pi f|}{V(x)} \\ & \leq & \sup_{x \in \mathcal{X}} \left(\Vert f \Vu + \frac{\pi V}{V(x)}\right)
 \leq  \Vert f \Vu + \frac{\pi(J)(K-\lambda)}{(1-\lambda) \inf_{x \in \mathcal{X}}V(x)}\\ & \leq & \Vert f \Vu + \frac{K-\lambda}{1-\lambda}.
\end{eqnarray*} 
\end{proof}

\begin{rem}
In many specific examples one can obtain (with some additional effort) sharper bounds for $\pi V,$ $\pi V^2,$ $\Vert \fc \Vu$ or at least bound $\pi(J)$ away from 1. However in general we assume that such bounds are not available and one will use Corollary \ref{cor: asVarBound}.
\end{rem}

\section{Example}\label{Sec: Example}

The simulation experiments described below are designed to compare
the bounds proved in this paper with actual errors of MCMC estimation. Assume that $y=(y_{1},\ldots,y_{t})$ is an i.i.d.\ sample from the normal distribution $\nor(\mu,\kappa^{-1})$, where $\kappa$ denotes the reciprocal  of the variance.
Thus we have 
\begin{equation*}
  \qquad p(y|\mu,\kappa)=p(y_{1},\ldots,y_{t}|\mu,\kappa)
     \propto {\kappa^{t/2}}\exp\left[-\frac{\kappa}{2}
         \sum_{j=1}^t(y_{j}-\mu)^{2}\right].
\end{equation*}
The pair $(\mu,\kappa)$ plays the role of unknown parameter. To make things simple, let us consider ``uninformative improper priors'' that is assume that
$p(\mu,\kappa)=p(\mu)p(\kappa)\propto \kappa^{-1}$. The posterior density is then 
\begin{equation*}
\begin{split}
  \qquad p(\mu,\kappa|y)&\propto p(y|\mu,\kappa)p(\mu,\kappa)\\
         &\propto \kappa^{t/2-1}\exp\left[-\frac{\kappa t}{2}
          \left(s^{2}+(\bar y-\mu)^{2}\right)\right],
\end{split}
\end{equation*}
where
\begin{equation*}
  \qquad \bar y=\frac{1}{t}\sum_{j=1}^{t}y_{j},
             \quad s^2=\frac{1}{t}\sum_{j=1}^{t}(y_{j}-\bar y)^{2}.
\end{equation*}
Note that $\bar y$ and $s^{2}$ only determine the location and scale of the posterior.
We will be using a Gibbs sampler, whose performance does not depend on scaling and location, therefore without loss of generality  we can assume that $\bar y=0$ and $s^{2}=t$.
Since $y=(y_{1},\ldots,y_{t})$ is kept fixed, let us slightly abuse notation 
by using symbols $p(\kappa|\mu)$, $p(\mu|\kappa)$ and $p(\mu)$ for 
$p(\kappa|\mu,y)$, $p(\mu|\kappa,y)$ and $p(\mu|y)$, respectively. 
Now, the Gibbs sampler
consists of drawing samples intermittently from both the conditionals. 
Start with some $(\mu_{0},\kappa_{0})$. Then, for $i=1,2,\ldots$,
\begin{itemize}
 \item $\kappa_{i}\sim \gam\left({t}/{2},({t}/{2})(s^{2}+\mu_{i-1}^{2})\right)$,
 \item $\mu_{i}\sim \nor\left(0,1/(\kappa_{i}t)\right)$.
\end{itemize}
If we are chiefly interested in $\mu$ then it is convenient to consider the two small steps 
$\mu_{i-1}\to \kappa_{i}\to \mu_{i}$ together. The transition density is
\begin{equation*}
\begin{split}
  \qquad p(\mu_{i}|\mu_{i-1})&=\int p(\mu_{i}|\kappa) p(\kappa|\mu_{i-1}) \d\kappa\\ 
         &\propto \int_{0}^{\infty}
           \kappa^{1/2}\exp\left[-\frac{\kappa t}{2}\mu_{i}^2\right] \times\\
         &\qquad\qquad \times \left(s^{2}+\mu_{i-1}^{2}\right)^{t/2} \kappa^{t/2-1}  
           \exp\left[-\frac{\kappa t}{2} \left(s^{2}+\mu_{i-1}^{2}\right)\right] \d\kappa\\
         &=  \left(s^{2}+\mu_{i-1}^{2}\right)^{t/2} \int_{0}^{\infty}  \kappa^{(t-1)/2}  
           \exp\left[-\frac{\kappa t}{2} \left(s^{2}+\mu_{i-1}^{2}+\mu_{i}^{2}\right)\right]
                                                                               \d\kappa\\
         &\propto \left(s^{2}+\mu_{i-1}^{2}\right)^{t/2}
               \left(s^{2}+\mu_{i-1}^{2}+\mu_{i}^{2}\right)^{-(t+1)/2}.
\end{split}
\end{equation*}
The proportionality constants concealed behind the  $\propto$ sign depend only on $t$.
Finally we fix scale letting $s^{2}=t$ and get
\begin{equation}\label{eq: trans}
  \qquad p(\mu_{i}|\mu_{i-1})\propto \left(1+\frac{\mu_{i-1}^{2}}{t}\right)^{t/2}
               \left(1+\frac{\mu_{i-1}^{2}}{t}+\frac{\mu_{i}^{2}}{t}\right)^{-(t+1)/2}.
\end{equation}
If we consider the RHS of \eqref{eq: trans} as a function of $\mu_{i}$ only, we can regard the first factor as constant and write 
\begin{equation*}
  \qquad p(\mu_{i}|\mu_{i-1})
\propto  \left(1+\left(1+\frac{\mu_{i-1}^{2}}{t}\right)^{-1}\frac{\mu_{i}^{2}}{t}\right)^{-(t+1)/2}.
\end{equation*}                                                                                 
It is clear that the conditional distribution of random variable 
\begin{equation}\label{eq: t}
  \qquad \mu_{i} \left(1+\frac{\mu_{i-1}^{2}}{t}\right)^{-1/2}
\end{equation}
is t-Student distribution with $t$ degrees of freedom.
Therefore, since the t-distribution has the second moment equal to $t/(t-2)$ for $t>2$, we infer that
\begin{equation*}
  \qquad \Ex(\mu_{i}^{2}|\mu_{i-1})
              =\frac{t+\mu_{i-1}^{2}}{t-2}.
\end{equation*}
Similar computation shows that the posterior marginal density of $\mu$ satisfies
\begin{equation*}
\qquad
p(\mu)\propto  \left(1+\frac{t-1}{t}\frac{\mu^{2}}{t-1}\right)^{-t/2}.
\end{equation*}
Thus the stationary distribution of our Gibbs sampler is rescaled t-Student with $t-1$ degrees of freedom. Consequently we have 
\begin{equation*}
  \qquad \Ex_{\pi} \mu^2=\frac{t}{t-3}.
\end{equation*}

\begin{prop}[Drift]\label{pr: Driftex}
Assume that $t\geq 4$. Let
\begin{equation*}
  \qquad  V^{2}(\mu):=\mu^2+1
\end{equation*}
and $J=[-a,a]$. The transition kernel of the (2-step) Gibbs sampler
satisfies
\begin{equation*} 
\qquad  PV^{2}(\mu) \leq \begin{cases}
                         \lambda^{2} V^{2}(\mu) & \text{ for } |\mu|>a;\\
                          K^{2} & \text{ for } |\mu|\leq a,\\
                        \end{cases}
\end{equation*} 
provided that $a>\sqrt{t/(t-3)}$. The quantities $\lambda$ and $K$ 
are given by
\begin{equation*}
\begin{split}
\qquad  \lambda^{2}&=\frac{1}{t-2}\left(\frac{2t-3}{1+a^{2}}+1\right),\\
         K^{2}&=2+\frac{a^{2}+2}{t-2}.
\end{split}
\end{equation*}
Moreover, 
\begin{equation*} 
\qquad  \pi (V^{2}) =\frac{2t-3}{t-3}.
\end{equation*} 
\end{prop}

\begin{proof}
It is enough to use the fact that
 \begin{equation*}
  \qquad PV^{2}(\mu)=\Ex(\mu_{i}^{2}+1|\mu_{i-1}=\mu)=
           \frac{t+\mu^{2}}{t-2}+1
\end{equation*}
and some simple algebra. Analogously,
$\pi(V^{2})= \Ex_{\pi} \mu^2+1$.
\end{proof}

\def\pmin{p_{\rm min}}

\begin{prop}[Minorization]\label{pr: Minex}
Let $\pmin$ be a subprobability density given by
\begin{equation*}
  \qquad \pmin(\mu)=
            \begin{cases}
                          p(\mu|a) & \text{ for } |\mu|\leq h(a);\\
                          p(\mu|0) & \text{ for } |\mu|> h(a),\\
                        \end{cases}
\end{equation*}
where $p(\cdot|\cdot)$ is the transition density given by \eqref{eq: trans} and
\begin{equation*}
  \qquad h(a)=
        \left\{a^{2}\left[\left(1+\frac{a^{2}}{t}\right)^{t/(t+1)}-1\right]^{-1}-t\right\}^{1/2}.
\end{equation*} 
Then $|\mu_{i-1}|\leq a$ implies  $p(\mu_{i}|\mu_{i-1})\geq \pmin(\mu_{i})$. Consequently,
if we take for $\nu$ the probability measure with the normalized density  $\pmin/\beta$ 
then the small set Assumption \ref{as: SmallSetCond} holds for $J=[-a,a]$. 
Constant $\beta$ is given by
\begin{equation*}
  \qquad \beta=1-\Pr\left(|\vartheta|\leq h(a)\right)
   +\Pr\left(|\vartheta|\leq \left(1+\frac{a^{2}}{t}\right)^{-1/2} h(a)\right),
\end{equation*} 
where $\vartheta$ is a random variable with t-Student distribution with $t$ degrees of freedom.
\end{prop}

\begin{proof}
The formula for $\pmin$ results from minimization of $p(\mu_{i}|\mu_{i-1})$ with respect
to $\mu_{i-1}\in[-a,a]$. We use \eqref{eq: trans}. 
First compute $(\d /\d \mu_{i-1}) p(\mu_{i}|\mu_{i-1})$ to check that the function has to attain minimum either at 0 or at $a$. Thus
\begin{equation*}
  \qquad \pmin(\mu)=
            \begin{cases}
                          p(\mu|a) & \text{ if }  p(\mu|a) \leq p(\mu|0) ;\\
                          p(\mu|0) & \text{ if }  p(\mu|a) > p(\mu|0).\\
            \end{cases}
\end{equation*}
Now it is enough to solve the inequality, say, $p(\mu|a) \leq p(\mu|0)$ with respect to
$\mu$. Elementary computation shows that this inequality is fulfiled iff $\mu\leq h(a)$. 
The formula for $\beta$ follows from \eqref{eq: t} and from the fact that 
\begin{equation*}
  \qquad \beta=\int \pmin(\mu)\d\mu=
        \int_{|\mu|\leq h(a)} p(\mu|a) \d\mu+\int_{|\mu|> h(a)} p(\mu|0)\d\mu.
\end{equation*}
\end{proof}

\begin{rem}
 It is interesting to compare the asymptotic behavior of the constants in
Propositions \ref{pr: Driftex} and \ref{pr: Minex}  for $a\to\infty$.
We can immediately see that $\lambda^{2}\to 1/(t-2)$ and $K^{2}\sim a^{2}/(t-2)$. Slightly more
tedious computation reveals that $h(a)\sim {\const}\cdot a^{1/(t+1)}$ and consequently 
$\beta\sim{\const}\cdot a^{-t/(t+1)}$. 
\end{rem}


The parameter of interest is the posterior mean (Bayes estimator of $\mu$). Thus we 
let $f(\mu)=\mu$ and $\th=\Ex_{\pi}\mu$. Note that our chain $\mu_{0},\ldots,\mu_{i},\ldots$
is a zero-mean martingale, so  $\fc=f$ and
\begin{equation*}
  \qquad \asvar=\Ex_{\pi}(f^{2})=\frac{t}{t-3}.
\end{equation*} 
Obviously we have $\Vert f\Vu =1$.

In the experiments described below, $t=50$ is kept fixed. Other
experiments (not reported here) show that the value of $t$ has little influence on the results. Table 1 illustrates inequalities in Theorem \ref{th: BasicMSE}
and Corollary \ref{cor: BasicMSE}. The actual values of the MSE of our estimator
and the mean overshoot, viz.
\begin{equation*}
\begin{split}
  \qquad \textrm{MSE}&:=\Ex\,(\hthsn-\th)^2,\\
         \textrm{OS}&:=\Ex\, T_{R(n)}- n,
\end{split}
\end{equation*} 
are computed empirically, using 10000 repetitions of the experiment. They can be compared with the bounds in \ref{th: BasicMSE} and \ref{cor: BasicMSE}, named henceforth 
\begin{equation*}
\begin{split}
\qquad
      \textrm{BoundMSE}&:=\frac{\asvar}{n}\left(1+\frac{\n0}{n}\right)\\
       \textrm{BoundOS}&:=\n0=\frac{\Ex \tau_{1}^{2}}{\Ex \tau_{1}}-1.
\end{split}
\end{equation*}
In these formulas, we use the {true value} of $\asvar$, for which we have an analytical 
expression. Also $m=\Ex \tau_{1}=\pi(J)\beta$ is computed exactly while $\Ex \tau_{1}^{2}$ is approximated via a separate (very long) series of simulations.
given  for two choices of the ``small set'' $J=[-a,a]$.
We also show values of $m$ (mean length of a regeneration cycle) and $\beta$ 
(probability of regeneration).

\bigskip
{\small
\begin{center}
\begin{tabular}{|c|c|c|c|c|c|c|c|}
\hline
$n$ & $a$ & MSE\phantom{${}^{A^{A^A}}$} & BoundMSE & OS\phantom{${}^{A^{A^A}}$}  & BoundOS & $m$ & $\beta$\\
\hline
10  &  & 0.1062 & 0.1087   & 0.1099 &   &  &  \\
\cline{1-1}\cline{3-5}
100 & 5 & 0.0105 & 0.0107   & 0.1037 & 0.2134  & 1.1072 & 0.9032 \\
\cline{1-1}\cline{3-5}
1000&  & 0.0011 & 0.0011   & 0.1073 &  &  &  \\
\hline
10  & & 0.0821 & 0.2247   & 5.4768 &  &  &  \\
\cline{1-1}\cline{3-5}
100 &100& 0.0102 & 0.0118   & 5.4871 & 11.1196 & 6.5043 & 0.1537 \\
\cline{1-1}\cline{3-5}
1000& & 0.0011 & 0.0011   & 5.4337 &  & &  \\
\hline
\end{tabular} 
\bigskip

 Table 1. Actual values of the MSE and mean overshoot vs. bounds \ref{th: BasicMSE} and \ref{cor: BasicMSE}
\end{center}
}

Table 1 clearly shows that the inequalities in Theorem \ref{th: BasicMSE} are quite sharp.
The bound on MSE, which is of primary interest, becomes almost exact for large $n$.
The bound on the mean overshoot, which can be used to estimate the cost of the algorithm, is also very satisfactory.

We now proceed to the inequalities proved in Section \ref{Sec: asVar}
under the drift condition, Assumption \ref{as: DriftCond}. 
The final bounds in Corollary \ref{cor: asVarBound} are expressed in terms of the computable drift/minorization parameters, that is $\lambda$, $K$ and $\beta$. We also examine how the tightness of the final bounds is influenced by replacing
the true value of $\pi V^2$ by its upper bound. To this end we compute the bounds given
in Theorem \ref{th: asVarBound}, using the knowledge of $\pi V^2$.
In our example we compute $\lambda$, $K$, $\beta$ and also $\pi V^2$ 
via Propositions \ref{pr: Driftex} and \ref{pr: Minex} for different choices of $J=[-a,a]$. Parameter $t=50$ is fixed.  

Figure 1 shows how the two bounds on $\asvar$ depend on $a$. The black line corresponds
to the bound of Corollary \ref{cor: asVarBound} (ii) which involves only $\lambda$, $K$ and $\beta$.  The grey line gives the bound 
of Theorem \ref{th: asVarBound} (ii) which assumes the knowledge of $\pi V^2$.
The best values of both bounds, equal to $7.19$ and $5.66$, correspond to  $a=3.93$ and $a=4.33$, respectively. The actual value of the asymptotic variance is
$\asvar=1.064$.

\begin{center}
\includegraphics[height=12cm, width=\textwidth]{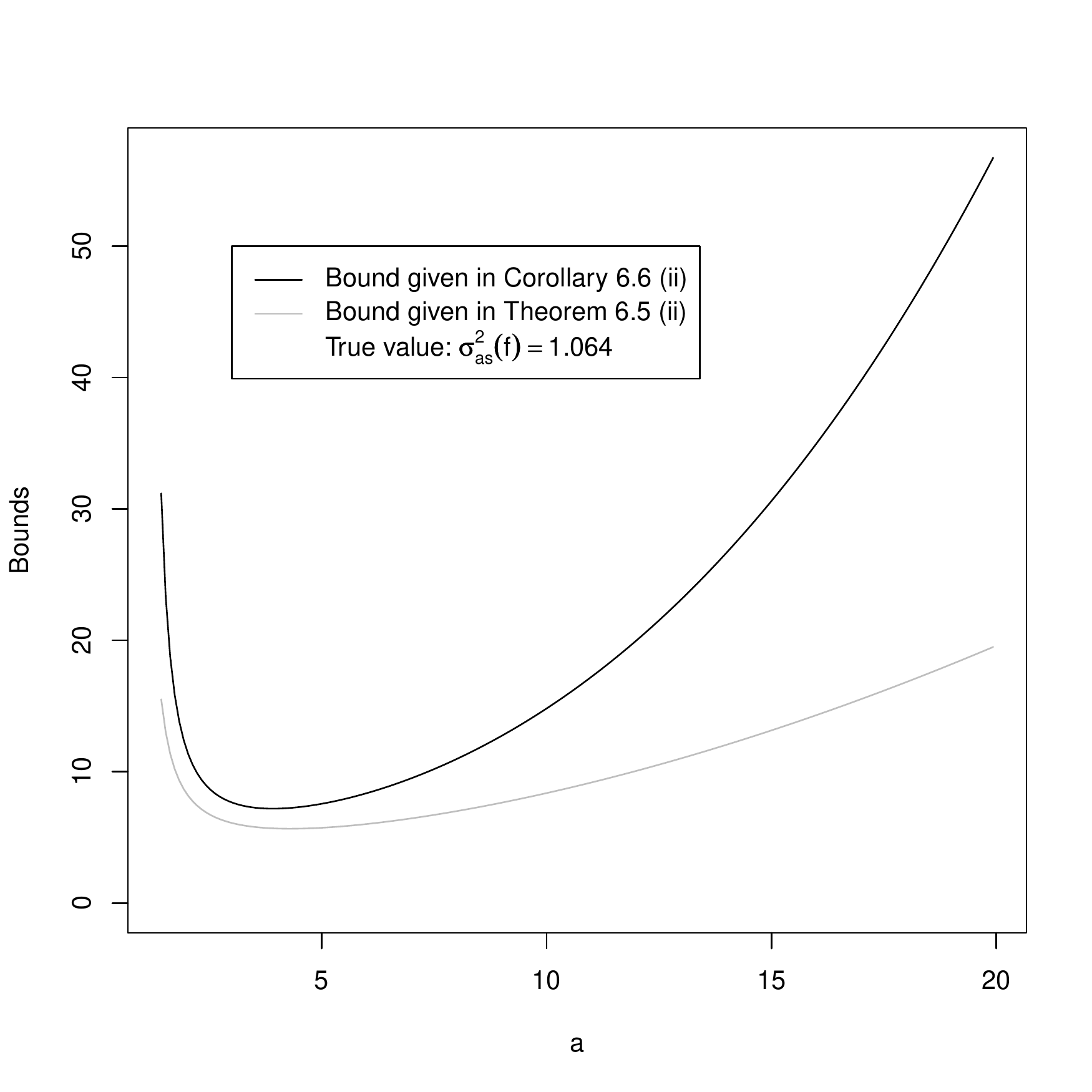}
Figure 1. 
Bounds for the asymptotic variance $\asvar$ as functions of $a$.
\end{center}

Figure 2 is analogous and shows two bounds on $\n0$.  Again, the black 
bound involves only the drift/minorization parameters while the grey one assumes
the knowlegde of $\pi V^2$.
The best bounds, $2.94$ and $2.50$, obtain for $a=4.73$ and $a=4.33$, respectively. 

\begin{center}
\includegraphics[height=12cm, width=\textwidth]{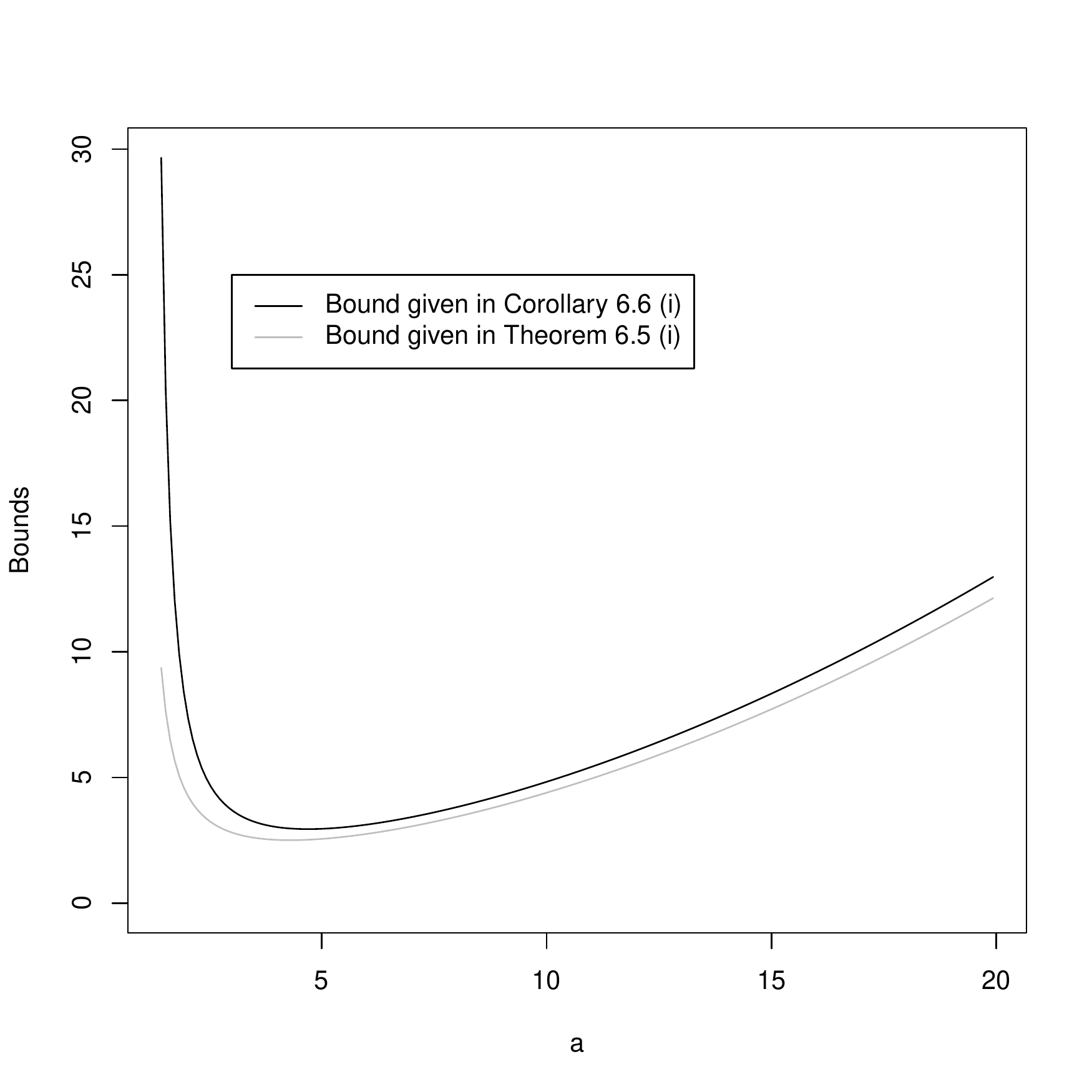}
Figure 2. Bounds for $\n0$ as functions of $a$.
\end{center}

In contrast with the inequalities of Theorem \ref{th: BasicMSE}, the bounds
of Theorem \ref{th: asVarBound} depend significantly on $t$, the size of
sample behind the posterior distribution. In Table 2 below we summarize
the best (with respect to $a$) bounds on $\asvar$ for three values of
$t$.

{\small
\begin{center}
\begin{tabular}{|c|c|c|c|c|}
\hline
$t$& $\asvar$& Bound \ref{th: asVarBound} (ii) & Bound \ref{cor: asVarBound} (ii)   \\
\hline
5  &      2.500    & 141.50  &  41.02    \\
\hline
50 &    1.064     &   7.19  & 5.66    \\
\hline
500&    1.006    &  4.33   & 3.99   \\
\hline
\end{tabular} 
\bigskip

 Table 2. Values of $\asvar$ vs. bounds \ref{th: asVarBound} and 
\ref{cor: asVarBound} for different values of $t$.
\end{center}
}

This clearly identifies the bottleneck of the approach: the bounds on $\asvar$ under drift condition in Theorem \ref{th: asVarBound} and Corollary \ref{cor: asVarBound} can vary widely in their sharpness in specific examples.  We conjecture that this may be the case in general for any bounds derived under drift conditions. Known bounds on the rate of convergence (e.g.\ in total variation norm) obtained under drift conditions are often very pessimistic, too (e.g. \cite{Bax, RoTw, JoHo}). However,  at present, drift conditions remain the main and most universal tool for proving computable bounds for Markov chains on continuous spaces. An alternative might be working with conductance but to the best of our knowledge, so far this approach has been applied successfully only to examples with compact state spaces (see e.g. \cite{Rud, MathNov} and references therein).

\section{Connections with other results}\label{Sec: Discuss} 

Our aim was to obtain nonasymptotic results concerning
the mean square error and confidence estimation in a possibly general setting relevant for MCMC applications in Bayesian statistics. We now discuss our results in context of related work.

\subsection{Related nonasymptotic results} A vast literature on nonasymptotic analysis of Markov chains is available in various settings. To place our results in this context we give a brief account, which by no means is extensive.
In the case of  finite state space, an approach based on the spectral decomposition was used in \cite{Ald, Gil, PerronLeon, MIS} to derive results of related type. For bounded functions and uniformly ergodic chains on a general state space, exponential
inequalities with explicit constants such as those in \cite{GlyOr, KontoLastraMeyn} can be applied to derive confidence bounds. Comparison of the required simulation effort for the same confidence interval (Sections \ref{Sec: Conf} and \ref{Sec:asV_unif}) shows that while exponential inequalities have sharper constants, our approach gives in this setting the optimal dependence on the regeneration rate $\beta$ and therefore can turn out more efficient in many practical examples.

Related results come also from studying concentration of measure phenomenon for dependent random variables. For the large body of work in this area see e.g. \cite{Marton}, \cite{Samson} and \cite{KoRa} (and references therein), where transportation inequalities or martingale approach have been used. These results, motivated in a more general setting, are valid for Lipschitz functions with respect to the Hamming metric. They also include expressions $\sup_{x,y \in \mathcal{X}} \|P^i(x, \cdot) - P^i(y, \cdot)\|_{\rm{tv}}$ and when applied to our setting, they are well suited for bounded functionals of uniformly ergodic Markov chains, but can not be applied to geometrically ergodic chains. For details we refer to the original papers and the discussion in Section 3.5 of \cite{Adamczak}.

For lazy reversible Markov chains, nonasymptotic mean square error bounds have been obtained for \emph{bounded} target functions in \cite{Rud} in a setting where explicit bounds on conductance are available. These results have been applied to approximating integrals over balls in $\Rl^d$ under some regularity conditions for the stationary measure, see \cite{Rud} for details. The Markov chains considered there are in fact uniformly ergodic, however in their problem when establishing (\ref{eqn_unif_erg}), $\beta$ turns out to be exponentially small and $\q > 1,$ hence conductance seems to be the natural approach to make the problem tractable in high dimensions.

Tail inequalities for \emph{bounded} functionals of Markov chains that are not uniformly ergodic were considered in \cite{Cle}, \cite{Adamczak} and \cite{DoGuMou} using regeneration techniques. These results apply e.g. to geometrically or subgeometrically ergodic Markov chains, however they also involve non-explicit constants or require tractability of moment conditions of random tours between regenerations. Computing explicit bounds from these results may be possible with additional work, but we do not pursue it here.

Tail inequalities for unbounded target function $f$ that can be applied to geometrically ergodic Markov chains have been established by Bertail and Cl\'emen\c con in \cite{BeCle3} by regenerative approach and using truncation arguments. However they involve non-explicit constants and can not be directly applied to confidence estimation.

Rates of convergence of geometrically ergodic Markov chains to their stationary distributions have been investigated in many papers. The typical setting is similar to our Section \ref{Sec: asVar}, i.e. one assumes a geometrical drift to a small set and a one step minorization condition. Moreover, to establish convergence rates one requires an additional condition that implies aperiodicity, which was not needed for our purposes. Most of the authors focus either on the total variation distance \cite{RobRos, Ros02, RoTw, JoHo, Ros95} or its weighted version \cite{Fort, Bax}. Such results, although of utmost theoretical importance, do not directly translate into bounds on the accuracy of estimation, because they allow us to control only the bias of estimates and the so-called burn-in time. Moreover we note, that in the drift condition setting \begin{itemize} \item convergence to stationarity is in fact not needed, we only require a bound on the asymptotic variance and on the overshoot, c.f. \mbox{Section \ref{Sec: asVar},} \item to obtain explicit convergence rates, some version of our Proposition \ref{pr: Bax} is always needed (c.f. for example Section 4 of \cite{Bax}) and it is in fact one of several steps required for the bound, whereas we are using Proposition \ref{pr: Bax} directly, avoiding other steps that could weaken the results.   \end{itemize}

\subsection{Nonasymptotic vs asymptotic confidence estimation} Since nonasymptotic analysis of complicated Markov chains appears difficult, practitioners often validate MCMC estimation by a convergence diagnostics (see e.g. \cite{CoCa, GeRu} and references therein). It is however well-known that this may lead to overoptimistic conclusions, stopping the simulation far to early, and introducing bias \cite{CoRoRo, Matt}. Designing asymptotic confidence intervals based on CLTs for Markov chains is often perceived as a reasonable trade-off between rigorous analysis of the algorithm and convergence heuristics and is referred to as honest MCMC estimation, c.f. \cite{Geyer, HoJo_honest}.

In what follows we argue that the nonasymptotic confidence estimation presented in the current paper requires verifying essentially the same assumptions as asymptotic confidence estimation. We also compare implementational difficulties. 

Asymptotic confidence estimation for Markov chains is done e.g. by establishing Edgeworth expansions (see \cite{BeCle1, BeCle2}) or by applying the Glynn and Whitt sequential procedure \cite{GlynnWhitt} in the Markov chain context. Both methods rely heavily on strongly consistent estimation of the asymptotic variance. There has been a lot of work done recently to analyse asymptotic variance estimators for Markov chains and enable strongly consistent estimation under tractable assumptions \cite{FleJo, CaHaJoNe, BeLa_asVar, HoJoPrRos, BeCle2, BeCle1}. In particular we note the following. 
\begin{itemize} \item The most commonly used regenerative estimators (see e.g. \cite{CaHaJoNe, HoJoPrRos, BeCle1, BeCle2}) are known to be strongly consistent for geometrically ergodic Markov chains that satisfy a one step minorization condition and an integrability condition $\Ex_{\pi}|f|^{2+\delta} < \infty$ (Proposition 1 of \cite{CaHaJoNe}). \item Similarly, the non-overlapping and overlapping batch means estimators (see e.g. \cite{CaHaJoNe, FleJo}) are known to be strongly consistent for geometrically ergodic Markov chains that satisfy a one step minorization condition and an integrability condition $\Ex_{\pi}|f|^{2+\delta} < \infty$ (Proposition 4 of \cite{BeLa_asVar} and Theorem 2 of \cite{FleJo} respectively). \item Spectral variance estimators are known to be strongly consistent for geometrically ergodic Markov chains that satisfy a one step minorization condition and an integrability condition $\Ex_{\pi}|f|^{4+\delta} < \infty$ (Theorem 1 of \cite{FleJo}). \end{itemize}
We note that geometrical ergodicity is typically established by a drift condition similar to the one used in Section \ref{Sec: asVar} and the one step minorization condition usually boils down to our Assumption \ref{as: SmallSetCond}. As for integrability conditions, the drift condition implies $\pi V^2 < \infty$ and we require $f^2 < V^2.$ Checking $E_{\pi}|f|^{2+\delta} < \infty$ will be typically done by ensuring $|f|^{2+\delta} < V^2$ and is therefore comparable, whereas the condition $E_{\pi}|f|^{4+\delta} < \infty$ for spectral variance estimation is clearly stronger.

On the algorithmic side, regenerative asymptotic variance estimators require identifying regenerations, exactly as we do, whereas batch means and spectral variance estimators do not require this.              


Therefore we conclude, that if regenerations are identifiable, the price for the rigorous, nonasymptotic result is only as high as the difference between $\asvar$ and its upper bounds e.g. those in Section \ref{Sec: asVar}.

\section*{Acknowledgements}
Discussions with Jacek Weso{\l}owski helped prepare an early version of this paper. The authors are also grateful to the anonymous referee for his/her constructive comments.

\bigskip
\goodbreak

\appendix

\section{Proof of Proposition \ref{pr: Bax}} \label{sec_appendix_prop}


\def\HA{H}
\def\HJ{\tilde{H}}
\def\HT{\bar{H}}
\def\s{S}

\begin{proof} Under \eqref{eq: Drift} and Assumption \ref{as: SmallSetCond} we are to establish
\begin{equation*}
\qquad           
      \Ex_x\sum_{n=1}^{T-1}V(X_n)\leq
             \frac{\lambda (V(x)-1)}{1-\lambda}+\frac{(K-\lambda)}{\beta(1-\lambda)}-1.
\end{equation*}
The idea is to decompose the sum into shorter blocks, such that each block ends at a visit to $J$.
Let $\s:=\s_{0}:=\min\{n\geq 0: X_{n}\in J\}$ and $\s_{j}:=\min\{n>\s_{j-1}: X_{n}\in J\}$
for $j=1,2,\ldots$. Introduce the following notations:
\begin{equation*}
\begin{split}
\qquad       
     &\HA(x):=\Ex_{x}\sum_{n=0}^{\s}V(X_{n}), \text{ for } x\in\X,\\
     &\HJ:=\sup_{x\in J} \Ex_{x} \left(\sum_{n=1}^{\s_{1}} V(X_{n})\Big|\Gamma_{0}=0\right)
         =\sup_{x\in J} \int Q(x,\dy)\HA(y).
\end{split}    
\end{equation*}
Note that $H(x)=V(x)$ for $x\in J$ and that $Q$ denotes the normalized ``residual kernel''. 

Let us first bound $\HA(x)$.
It is easy to check that under \eqref{eq: Drift}, for every initial distribution,
$V(X_{n\land \s})/\lambda^{n\land \s}$ for $n=0,1,\ldots$ is a supermartingale with respect to
$\F_{n}:=\sigma(X_{0},\ldots,X_{n})$. Therefore
$\Ex_{x}V(X_{n\land \s})/\lambda^{n\land \s}\leq V(x)$ for every $x\in\X$ and $n=0,1,\ldots$. This inequality can be  multiplied by $\lambda^n$ and rewiritten as follows:
\begin{equation*}
\qquad           
      \Ex_{x}V(X_{\s})\lambda^{n-\s}\Ind(\s<n) + \Ex_{x}V(X_{n})\Ind(n\leq \s)
  		\leq\lambda^{n}V(x).
\end{equation*}
Now take a sum over $n=0,1,\ldots$ to obtain
\begin{equation*}\label{eq: superMG}
\qquad           
      \Ex_{x}V(X_{\s})\sum_{n=\s+1}^{\infty}\lambda^{n-\s}+\Ex_{x}\sum_{n=0}^{\s}V(X_{n})
          		\leq V(x)\sum_{n=0}^{\infty}\lambda^{n}
\end{equation*}
or, equivalently,
\begin{equation}\label{eq: superMG}
\qquad           
      \Ex_{x}V(X_{\s})\frac{\lambda}{1-\lambda}+H(x)
          		\leq V(x)\frac{1}{1-\lambda}.
\end{equation}
Consequently, since $\Ex_{x}V(X_{\s})\geq 1$, we have for exery $x$,
\begin{equation}\label{eq: atomic}
\qquad           
     H(x)\leq\frac{V(x)-\lambda}{1-\lambda}.
\end{equation}

From \eqref{eq: Drift} we obtain $PV(x)=(1-\beta)QV(x)+\beta \nu V\leq K$ for $x\in J$, so
$QV(x)\leq (K-\beta)/(1-\beta)$ and, taking into account \eqref{eq: atomic},
\begin{equation}\label{eq: HJ}
\qquad           
      \HJ\leq 
         \frac{(K-\beta)/(1-\beta)-\lambda}{1-\lambda}
         =\frac{K-\lambda-\beta(1-\lambda)}{(1-\lambda)(1-\beta)}.
\end{equation}
Recall that $T:=\min\{n\geq 1: \Gamma_{n-1}=1\}$. For $x\in J$ we thus have
\begin{equation*}
\begin{split}
\qquad           
      &\Ex_x\sum_{n=1}^{T-1}V(X_n)\\
          &=\Ex_x \sum_{j=1}^{\infty}\sum_{n=\s_{j-1}+1}^{\s_{j}} V(X_{n})
          \Ind(\Gamma_{\s_{0}}=\cdots=\Gamma_{\s_{j-1}}=0)\\
          &=\Ex_x \sum_{j=1}^{\infty}\Ex\left(\sum_{n=\s_{j-1}+1}^{\s_{j}} V(X_{n})\Bigg\vert
                \Gamma_{\s_{0}}=\cdots=\Gamma_{\s_{j-1}}=0\right)(1-\beta)^{j}\\
          &\leq \sum_{j=1}^{\infty} \HJ (1-\beta)^{j}\leq 
         \frac{K-\lambda}{\beta(1-\lambda)}-1,
\end{split}
\end{equation*}
by  \eqref{eq: HJ}.  For $x\not\in J$ we have to add one more term at the beginning:
\begin{equation*}
\begin{split}
\qquad           
      &\Ex_x\sum_{n=1}^{T-1}V(X_n)=\Ex_x \sum_{n=1}^{\s_{0}} V(X_{n})\\
          &+\Ex_x \sum_{j=1}^{\infty}\sum_{n=\s_{j-1}+1}^{\s_{j}} V(X_{n})
               \Ind(\Gamma_{\s_{0}}=\cdots=\Gamma_{\s_{j-1}}=0).
\end{split}
\end{equation*}
This extra term is equal to $\HA(x)-V(x)$ and we can use \eqref{eq: atomic} to bound it.
Finally we obtain
\begin{equation}\label{eq: final}
\qquad           
      \Ex_x \sum_{n=1}^{T-1}V(X_n)\leq \frac{\lambda(V(x)-1)}{1-\lambda}\Ind(x\not\in J)+
               \frac{K-\lambda}{\beta(1-\lambda)}-1. 
\end{equation}

\end{proof}

\end{document}